\newcommand{\CLASSINPUTtoptextmargin}{0.75in}%
\newcommand{\CLASSINPUTbottomtextmargin}{1in}%
\newcommand{\CLASSINPUTinnersidemargin}{0.63in}%
\newcommand{\CLASSINPUToutersidemargin}{0.63in}%
\documentclass[conference,10pt,letterpaper]{IEEEtran}%
\usepackage[hyphens]{url}
\usepackage{hyperref}
\hypersetup{breaklinks=true}

\usepackage{cite}
\usepackage{amsmath}
\usepackage{graphicx}
\usepackage{multirow}
\usepackage[none]{hyphenat}
\usepackage{float}
\usepackage{subfig}
\usepackage{dblfloatfix}
\usepackage{t1enc}
\usepackage{times}
\makeatletter

\def\@IMSauthorblockNAMEstyle{\normalfont\IMSauthorsize}
\def\@IMSauthorblockAFFILstyle{\normalfont\IMSaffilsize}
\def\@IMSauthorblockEMAILstyle{\normalfont\IMSaffilsize}
\def\IMSauthorblockNAME#1{%
\relax\@IMSauthorblockNAMEstyle%
#1%
}%
\def\IMSauthorblockAFFIL#1{%
\relax\@IMSauthorblockAFFILstyle%
\vskip\@IEEEauthorblockAtopspace
#1%
}%
\def\IMSauthorblockEMAIL#1{%
\relax\@IMSauthorblockEMAILstyle%
\vskip\@IEEEauthorblockAtopspace
#1%
}%
\newif\ifIsBlindReviewVersion
\def\IMSthispaperforblindreview{\IsBlindReviewVersiontrue}
\def\IMSthispaperforfinalpublication{\IsBlindReviewVersionfalse}
\def\@maketitle{\newpage
\bgroup\par\addvspace{0.5\baselineskip}\centering%
\ifCLASSOPTIONtechnote
   {\bfseries\large\@IEEEcompsoconly{\sffamily}\@title\par}\vskip 1.3em{\lineskip .5em\@IEEEcompsoconly{\sffamily}\@author
   \@IEEEspecialpapernotice\par{\@IEEEcompsoconly{\vskip 1.5em\relax
   \@IEEEtitleabstractindextextbox{\@IEEEtitleabstractindextext}\par
   \hfill\@IEEEcompsocdiamondline\hfill\hbox{}\par}}}\relax
\else
   \vskip0.2em{\IMStitlesize\ifCLASSOPTIONtransmag\bfseries\LARGE\fi\@IEEEcompsoconly{\sffamily}\@IEEEcompsocconfonly{\normalfont\normalsize\vskip 2\@IEEEnormalsizeunitybaselineskip
   \bfseries\Large}\@title\par}\vskip1.0em\par
   \ifCLASSOPTIONconference%
      {\@IEEEspecialpapernotice\mbox{}\vskip\@IEEEauthorblockconfadjspace%
       \mbox{}\hfill\begin{@IEEEauthorhalign}\@author\end{@IEEEauthorhalign}\hfill\mbox{}\par}\relax
   \else
      \ifCLASSOPTIONpeerreviewca
         {\@IEEEcompsoconly{\sffamily}\@IEEEspecialpapernotice\mbox{}\vskip\@IEEEauthorblockconfadjspace%
          \mbox{}\hfill\begin{@IEEEauthorhalign}\@author\end{@IEEEauthorhalign}\hfill\mbox{}\par
          {\@IEEEcompsoconly{\vskip 1.5em\relax
           \@IEEEtitleabstractindextextbox{\@IEEEtitleabstractindextext}\par\hfill
           \@IEEEcompsocdiamondline\hfill\hbox{}\par}}}\relax
      \else
         \ifCLASSOPTIONtransmag
           {\@IEEEspecialpapernotice\mbox{}\vskip\@IEEEauthorblockconfadjspace%
            \mbox{}\hfill\begin{@IEEEauthorhalign}\@author\end{@IEEEauthorhalign}\hfill\mbox{}\par
           {\vspace{0.5\baselineskip}\relax\@IEEEtitleabstractindextextbox{\@IEEEtitleabstractindextext}\vspace{-1\baselineskip}\par}}\relax
         \else
           {\lineskip.5em\@IEEEcompsoconly{\sffamily}\sublargesize\@author\@IEEEspecialpapernotice\par
           {\@IEEEcompsoconly{\vskip 1.5em\relax
            \@IEEEtitleabstractindextextbox{\@IEEEtitleabstractindextext}\par\hfill
            \@IEEEcompsocdiamondline\hfill\hbox{}\par}}}\relax
         \fi
      \fi
   \fi
\fi\par\addvspace{0.0\baselineskip}\egroup}

\def\IMStitlesize{\@setfontsize{\IMStitlesize}{18}{21pt}}
\def\IMSauthorsize{\@setfontsize{\IMSauthorsize}{12}{13pt}}
\def\IMSaffilsize{\@setfontsize{\IMSaffilsize}{12}{13pt}}
\def\IMScaptionsize{\@setfontsize{\IMScaptionsize}{8}{9pt}}
\def\IMSbibsize{\@setfontsize{\IMSbibsize}{8}{9pt}}

\def\@IEEEauthorblockNstyle{\IMSauthorsize\@IEEEcompsocnotconfonly{\sffamily}\@IEEEcompsocconfonly{\large}}
\def\@IEEEauthorblockAstyle{\IMSaffilsize\@IEEEcompsocnotconfonly{\sffamily}\@IEEEcompsocconfonly{\itshape}\@IEEEcompsocconfonly{\large}}
\def\@IEEEauthordefaulttextstyle{\IMSauthorsize\@IEEEcompsocnotconfonly{\sffamily}\sublargesize}

\def\thebibliography#1{\section*{\refname}%
    \addcontentsline{toc}{section}{\refname}%
    \IMSbibsize\@IEEEcompsocconfonly{\small}\vskip 0.3\baselineskip plus 0.1\baselineskip minus 0.1\baselineskip
    \list{\@biblabel{\@arabic\c@enumiv}}%
    {\settowidth\labelwidth{\@biblabel{#1}}%
    \leftmargin\labelwidth
    \advance\leftmargin\labelsep\relax
    \itemsep \IEEEbibitemsep\relax
    \usecounter{enumiv}%
    \let\p@enumiv\@empty
    \renewcommand\theenumiv{\@arabic\c@enumiv}}%
    \let\@IEEElatexbibitem\bibitem%
    \def\bibitem{\@IEEEbibitemprefix\@IEEElatexbibitem}%
\def\newblock{\hskip .11em plus .33em minus .07em}%
\ifCLASSOPTIONtechnote\sloppy\clubpenalty4000\widowpenalty4000\interlinepenalty100%
\else\sloppy\clubpenalty4000\widowpenalty4000\interlinepenalty500\fi%
    \sfcode`\.=1000\relax}

\long\def\@makecaption#1#2{%
\ifx\@captype\@IEEEtablestring%
\par\@IEEEtabletopskipstrut
\else
\@IEEEfigurecaptionsepspace
\fi
\setbox\@tempboxa\hbox{\normalfont\IMScaptionsize {#1.}\nobreakspace\nobreakspace #2}%
\ifdim \wd\@tempboxa >\hsize%
\setbox\@tempboxa\hbox{\normalfont\IMScaptionsize {#1.}\nobreakspace\nobreakspace}%
\parbox[t]{\hsize}{\normalfont\IMScaptionsize\noindent\unhbox\@tempboxa#2}%
\else
\ifCLASSOPTIONconference \hbox to\hsize{\normalfont\IMScaptionsize\hfil\box\@tempboxa\hfil}%
\else \hbox to\hsize{\normalfont\IMScaptionsize\box\@tempboxa\hfil}%
\fi\fi
\ifx\@captype\@IEEEtablestring%
\@IEEEtablecaptionsepspace
\else
\fi}

\newlength\tablecaptiontotableskip
\newlength\figuretocaptionskip
\def\@IEEEfigurecaptionsepspace{\vskip\figuretocaptionskip\relax}%
\def\@IEEEtablecaptionsepspace{\vskip\tablecaptiontotableskip\relax}%

\def\abstract{\normalfont%
\@IEEEabskeysecsize\bfseries\textit{\abstractname}\,\bfseries\textit{---}\,%
\@IEEEgobbleleadPARNLSP}%

\def\IEEEkeywords{\normalfont%
\@IEEEabskeysecsize\bfseries\textit{\IEEEkeywordsname}\,\bfseries\textit{---}\,%
\@IEEEgobbleleadPARNLSP}%
\def\endIEEEkeywords{\relax\vspace{0.67ex}%
\par\if@twocolumn\else\endquotation\fi%
\normalsize\normalfont}%

\def\@IEEEauthorblockNtopspace{0ex}
\def\@IEEEauthorblockAtopspace{1mm}
\def\tablename{Table}
\def\thetable{\arabic{table}}
\def\IEEEkeywordsname{Keywords}
\def\subsubsection{\@startsection{subsubsection}{3}{\z@}{1.5ex plus 1.5ex minus 0.5ex}%
{0.7ex plus .5ex minus 0ex}{\normalfont\normalsize\itshape}}%
\def\@seccntformat#1{\csname the#1dis\endcsname\relax}
\def\thesectiondis{\thesection.\hskip 0.5em}
\def\thesubsectiondis{{\hbox to\parindent{\Alph{subsection}.}}}
\def\thesubsubsectiondis{{\hbox to \parindent{\arabic{subsubsection})}}}
\def\theparagraphdis{{\hbox to \parindent{\alph{paragraph})}}}
\IEEEilabelindentA \parindent
\IEEEilabelindent \IEEEilabelindentA
\IEEEelabelindent \parindent
\IEEEdlabelindent \parindent
\IEEElabelindent \parindent

\newlength\@IMSparindent
\newcommand\IMSdisplayacksection[1]{%
\ifIsBlindReviewVersion%
\noindent\phantom{\parbox[t]{\columnwidth}{\normalbaselines\setlength{\parindent}{\@IMSparindent}{#1}\strut}}
\else%
\noindent\parbox[t]{\columnwidth}{\normalbaselines\setlength{\parindent}{\@IMSparindent}{#1}\strut}%
\fi%
}%

\makeatother

\begin{document}
\raggedbottom
%
%
%
\title{Exploring the effectiveness of documentary film for science communication}
%
%
%
\IMSthispaperforblindreview
\IMSthispaperforfinalpublication
\author{\IEEEauthorblockN{\hspace{5mm} Sunanda Prabhu Gaunkar \hspace{5mm}}
\IEEEauthorblockA{
\textit{University of Chicago}\\
Chicago, IL, USA \\
spg@uchicago.edu}
\and
\IEEEauthorblockN{\hspace{5mm} Ellen Askey \hspace{5mm}}
\IEEEauthorblockA{
\textit{University of Chicago}\\
Chicago, IL, USA \\
ellen.askey@gmail.com}

\and
\IEEEauthorblockN{\hspace{5mm} Meira Chasman \hspace{5mm}}
\IEEEauthorblockA{
\textit{University of Chicago}\\
Chicago, IL, USA \\
mchasman@uchicago.edu}
\and
\IEEEauthorblockN{\hspace{5mm} Koksuke Takaira \hspace{5mm}}
\IEEEauthorblockA{
\textit{University of Chicago}\\
Chicago, IL, USA \\
ktakaira@uchicago.edu}
\and
\IEEEauthorblockN{\hspace{5mm} Calahan Smith \hspace{5mm}}
\IEEEauthorblockA{
\textit{University of Chicago}\\
Chicago, IL, USA \\
calahansmith@uchicago.edu}
\and
\IEEEauthorblockN{\hspace{5mm} Amanda Murphy\hspace{5mm}}
\IEEEauthorblockA{
\textit{University of Chicago}\\
Chicago, IL, USA \\
akmurphak@uchicago.edu}
\and
\IEEEauthorblockN{\hspace{5mm} Nancy Kawalek \hspace{5mm}}
\IEEEauthorblockA{
\textit{University of Chicago}\\
Chicago, IL, USA \\
kawalek@uchicago.edu}
}
\maketitle
%
%
%
\begin{abstract}
The complexity of science and its frequent lack of accessibility often creates disinterest among the general public. Furthermore, there exists a gap between the public perception of science and the reality of scientific research severely limits the scope of public engagement with science. This growing disparity is exacerbated by existing popular media that invariably portray scientists as either geniuses who constantly have breakthroughs or nerds without social skills buried in books. A  new docuseries, \emph{Curiosity: The Making of a Scientist},  created at the STAGE Lab\cite{stage} at the University of Chicago, addresses this issue by increasing the awareness and excitement around both science and scientists. Each film of  \emph{Curiosity} focuses on a single scientist and tells the story of how they became a scientist by interweaving elements of their personal life together with the successes and failures they encounter in their scientific work. One of the main  goals of the series is to present the stories of the people behind the science so that the public can create a personal connection to the scientists as a way to welcome the audience into the world of science. Our intended audience includes scientists and also non-scientists with little or no previous exposure to quantum physics or even science.
The pilot of the series, SUPERPOSITION\cite{superposition}, is a 25-minute film about a graduate student in quantum physics at the University of Chicago. To evaluate the success of this film, it was screened to several audiences covering most age-groups and audience members were requested to fill a detailed anonymous survey \cite{survey}after the viewing. Via these surveys, viewers reported an increased interest in science and a connection to the personal story of the graduate student that was purposefully woven throughout the film.  Survey responses show this film stimulated the audience’s interest in and appreciation for science. Further analysis of the survey results also suggests creative documentary films are an effective means of science communication that can bring scientific research along with the human enterprise of science to a broad audience. By inviting viewers to discover the world of science via universal themes that resonate with their lives, regardless of their familiarity with or affinity for science, \emph{Curiosity} intends to educate the public about the world of science, scientists, and the realities of scientific research, thereby increase public engagement with science.   

\end{abstract}
\begin{IEEEkeywords}
story, science communication, documentary.
\end{IEEEkeywords}

\section{Introduction}

In order to address the inaccessibility of science amongst the general public, a collaborative and interdisciplinary team at The University of Chicago’s STAGE Lab is developing a documentary-style web series uncovering scientists’ lives, motivations, and thinking. This docuseries titled, \emph{Curiosity: The Making of a Scientist}is made for a broad audience, demystifying science and dispelling stereotypes. Its first film, SUPERPOSITION, follows young quantum physicist Nate Earnest-Noble through the final year of his PhD at The University of Chicago \cite{schuster}. 

The team chose Nate Earnest-Noble because as the subject for SUPERPOSITION because his research topic had a close metaphor to his personal life – a humanizing connection that had the potential to make Nate’s scientific research relatable to someone who might have never heard of quantum physics. As Nate struggles to build a device that supports quantum superposition, we see that his personal life mirrors this state of being, as he is caught between his logical and emotional sides. The film was structured around emotional events in Nate’s everyday life, especially the people who influence his behaviors and decisions.  We draw parallels between the scientist’s personal life and work, highlighting the choices that guide their career. With SUPERPOSITION, we intend to familiarize a broad audience with a researcher’s life in quantum science and engineering. The film also educates the audience about what is involved in being a grad student, which is very valuable to young audiences who are considering a career track in this field.

There exists a gap between the reality of scientific research and the public perception of it \cite{perception}, which greatly influences how much the public will be willing to support science. We are all in the midst of an exciting quantum revolution that will impact everyone. People and governments need to make crucial decisions surrounding financial spending, ethics, and keeping up with or even spearheading the world’s technological advances. It is vital to get people to embrace the new technological advances that quantum science will allow and raise awareness to facilitate the wide public discussion that those “quantum leaps” will require. Without major efforts to engage and educate the public, a large fraction of society will remain disenfranchised and disconnected from this revolution, crucially impeding both public legitimacy and proper workforce in this important field. Because quantum physics is primarily taught at an undergraduate level in physics curriculum, the concepts may be too elusive and abstract for the broader public to learn independently. This difficulty is compounded by the limited availability of easily understandable resources and teaching materials that do not employ scientific jargon and equations. There is a long  line  of  art  and  media projects that aim to educate the general public on different scientific concepts, including but not limited to PBS’s NOVA\cite{nova}, Bill Nye the Science Guy \cite{nye} (and plenty of other children’s programming), many science-based internet series, and documentaries such as Explained\cite{explained}. However, the \emph{Curiosity}  docuseries, and in particular, SUPERPOSITION differs from these approaches because it attempts to make quantum science approachable by focusing on the everyday life of the scientist, including the scientist’s journey, personal life outside of the lab, and each success and failure through their scientific endeavors. SUPERPOSITION aim to bridge this gap by bringing viewers into the labs and lives of a young quantum scientist, and explains quantum computing in a way that can be easily understood through Nate’s story getting his PhD.  In this sense, the primary goal is to make the scientific process less opaque and more emotionally relatable, while simultaneously communicating scientific content.

In the following sections, we describe the concept of the film SUPERPOSITION through the surveys submitted by audience members. We intend to explain our motives in making the film, as well as what aspects were successful in receiving the intended audience response. We intend for our analysis of the surveys to encourage more scientists to collaborate with filmmakers and artists. This paper aims to outline how films in this style can benefit the scientific community and drive further interest in the sciences.

\section{Methods}

SUPERPOSITION was completed in June 2020, and then screened for private audiences and submitted to film festivals. Since then, 600 people have watched the documentary film, and 156 people have responded to a survey about the film after the viewing. The survey included15 questions, some demographic, while others attempted to obtain a detailed look at how this film impacted the viewer. 

The team targeted several different audiences while screening SUPERPOSITION intending to achieve a variety of goals set forth by this project. For curious adults, eager to connect or reconnect with science, we hoped to make them aware of different fields of research and the cutting edge work being pursued within, and also gives them a glimpse of the scientific thought process in action, in and out of the lab. For a young audience, the film offered an honest picture of what it means to be a scientist, along with a bountiful menu of scientific endeavors from which to choose. For audience members within the scientific community, SUPERPOSITION hoped to inspire scientists to become more committed and engaged in communicating their research – in a straightforward, understandable manner – to a diverse audience. For audiences with lesser affinity to science, and suspicious of science, we hoped to present the reality of the scientific enterprise to dispel stereotypes and demystify science. 

The survey functioned to confirm whether our goals of driving scientific interest and connecting with diverse audiences were met. The survey also functioned to pinpoint the strongest and weakest aspects of the documentary to carry those lessons to future films. Survey questions were provided in an online format through Google Forms, wherein some questions were multiple choice and others required qualitative, short answer responses.

 \section{Findings}
 
Because this documentary is primarily focused on the scientist’s character rather than simply the content of his work, we hoped viewers would find it easier to relate to and understand the mindset     of a scientist. We, therefore, hoped to foster in our viewers an overall positive, more emotional disposition towards the stark idea of science.
 
\begin{figure}
\centering
\includegraphics[width=75mm]{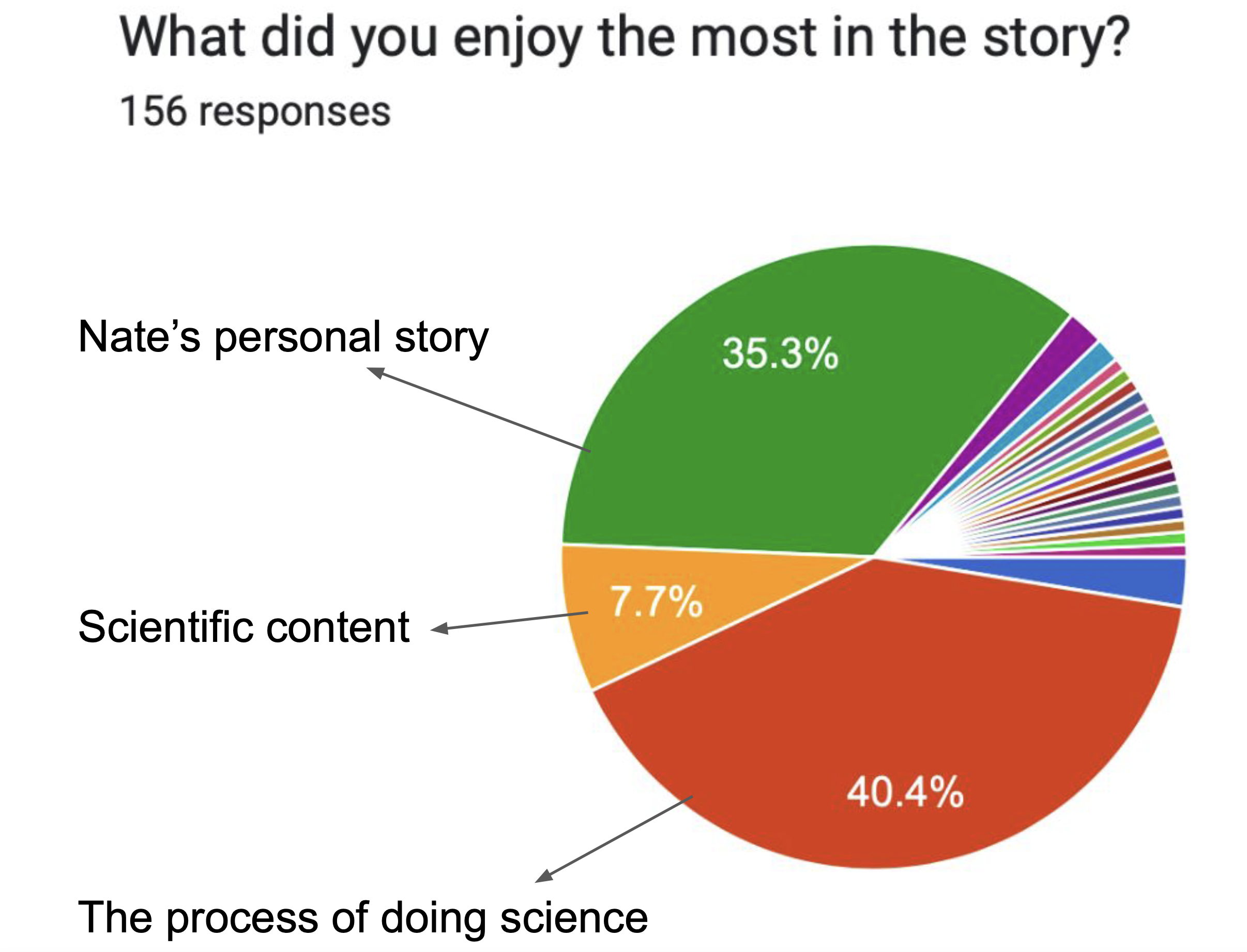}
\caption{Audience survey response to the question, "What did you enjoy the most in the story?"}
\label{fig:figure1}
\end{figure}

Survey responses to some questions (Figure.~\ref{fig:figure1}) were split: when asked “What did you enjoy most in the story?”, 35.5 \% answered “Nate’s personal story,” and 40 \% answered “the process of doing science.” These percentages align with the fact that while the film does explain scientific concepts, it portrays the scientist himself more than his  work and draws metaphors between the scientist’s work and personal life, and this structure was found enjoyable by audiences. \emph{Curiosity}’s main goal is to communicate science authentically and bring the compelling, relatable aspects of science such as the scientist’s personal journey and the scientific process to a broad audience, and the audience’s response supports this ideal. 

\textbf{The Scientific Process}. ~Audience members were also given the opportunity to offer longer, more personal answers on questions such as “Did you learn something new about the process or world of science?” A primary takeaway was that being a scientist and needing to engage in this research process was incredibly difficult. One responder said that they “got a real sense of the importance of accepting failure and frustration as a part of the scientific process.” Another response was “I enjoyed how honest everyone was on the process, acknowledging that a majority of the time will be spent frustratedly searching for a solution and that’s ok.” The audience’s takeaways focused more on the scientific process and what it means on an emotional level to sustain years of hard work, failure, and rejection, with dedication and perseverance. These responses suggest that the episode helped educate the audience about the scientific process and the emotional aspects involved in a scientist’s life, that are inherent to the research process, such as the importance of accepting failure and frustration. Certain demographics such as young students who are considering pursuing a PhD, young scientists who are pursuing their own struggles, and close acquaintances of quantum scientists who only had a bird’s eye view of their lives, mentioned that they particularly benefitted from knowing about this aspect of a scientist’s life. 

\textbf{Lives of scientists}. ~Some audience members were surprised by the unexpected drama and professional pressures of a graduate student’s life. For example, Nate Earnest-Noble, the scientists in SUPERPOSITION, is confronted with a dilemma of whether he can publish his work when a competing lab releases their findings. One audience member remarked, “I was struck by how ‘paper driven’ Nate’s inquiry seemed to be. I know the story was about a student but the search for scientific results seemed to be trumped by the need to get something into print. This film was the making of a scientist, so it was appropriate, and I found it curious. It seemed that his three years of work would have been without purpose had he not published. Is it so that he would not have gotten his doctorate if someone else’s research ‘beat’ him to the press?” The scientific process does not only occur in a lab– it involves writing papers for journals, applying for grants, meeting with advisors. The documentary exposes these elements of a scientific career that are simultaneously stressful and mundane, a sentiment that is relatable for workers in all sorts of industries. This workplace drama successfully uncovers the often overlooked personal aspect of the scientific process. 

\textbf{Communicating scientific concepts}. ~Other audience members remarked that they also learned new scientific concepts in a way that excited them and gained insight into Nate’s work: “I learned more about qubits and how Nate was trying to achieve superposition” and “it was fascinating to see how the little parts of future quantum computers are made.” These responses indicate that these viewers were able to grasp a basic understanding of concepts in quantum physics that were shown in the film. In regular pedagogical learning, due to it's inherently abstract and mathematical nature, quantum physics requires several prerequisites, such as a foundation in the sciences and mathematics, and is primarily taught at the undergraduate and higher levels. But the survey responses asserted that we could communicate the concepts of quantum physics in a way that is comprehensible and accessible to the broader public.

Furthermore, when asked they thought about the amount of scientific content in the story, 114 members of the audience felt that the amount of scientific content was just right, 1 member felt that it was excessive, and 39 members felt that it was too little. These survey responses suggest that we were able to provide adequate scientific content in the story in a way that did not away from the film’s entertainment value. This data will guide the editing team for future episodes in balancing the amount of content related to the scientist’s personal life with the content about their research to maximize positive reception.

\textbf{Building metaphors}. ~Several audience members grasped Nate’s need to embrace his logical and emotional sides. Nate's research is to build a device that lives in “quantum superposition,” which means that it can be in two states – “on” and “off” – at the same time. This concept aligns with Nate's state of mind. He struggles to balance his emotional and logical sides, which he calls "the two states of Nate." In this way, the viewer becomes acquainted with conceptually abstract ideas and complex scientific topics by relating them to the scientist’s personal journey. This metaphor resonated well with audiences. One response said: “I learned things I didn’t know previously about the world of quantum computing!  It was also comforting to see these scientists as human beings with personal stories especially as I’m taking my next steps into a PhD program.” Another remarked: “I personally enjoyed the emotional side of working in science, how it’s a balancing act: both accepting that you are a human, but also not mixing too much emotion with daily scientific practice.” In illustrating the research content with this intimate lens, Curiosity invites viewers to discover the world of science through the little-told stories of those in the lab. This balance, or paradox, was communicated effectively in the film– some viewers appreciated the complexity of reconciling the colder, objective sterility of science with the subtle, subjective nature of humanity.

\textbf{Appreciation for artistic expressions of science content}. ~Finally, an unexpected bit of feedback came from an anonymous science professor, who answered: “In particular, I learned about how you and your team view it...wonderful process, team and perspective. I look forward to seeing more of your work, and will recommend it to others and use it in my classes. . .” In this case, the scientist gained insight into the artistic process of filmmaking through this documentary. Therefore, an interdisciplinary project such as SUPERPOSITION can excite scientists about art, leading to more scientific content having artistic representations. More broadly, science documentaries can serve as a bridge between these two fields of study. STAGE fosters collaboration between artists and scientists for this very reason. An interdisciplinary team of from diverse disciplines such as Anthropology, Art History, Biology, Cinema, and Media Studies, English, Geography, Mathematics, Molecular Engineering, Music, Neuroscience, Philosophy, Physics, and Public Policy, created SUPERPOSITION. Both scientists and artists help create scientific content and craft narrative content. The team provided the project with numerous points of view, guiding story development to engage a diverse audience. 

Our survey results show that this method of scientific communication via documentary inspires a greater interest in scientific fields. Even if the viewer does fully understand the concepts present within the film, the fact that they care about the success or failure of a scientist’s pursuits is an incredibly important first step in driving people to become more involved in the sciences. Films like SUPERPOSITION are not the flat, classroom presentations. Rather, they serve as an audience member’s gateway into the larger scientific world, dramatically lowering the sciences’ barrier to entry. The \emph{Curiosity} project helps make the pursuit of a scientific career seems not only attainable, but worthwhile.

 \section{Future Steps}

Going forward, we will continue to emphasize the personal stories behind the scientists we are filming. The films in the series will showcase a diverse and inclusive group of scientists in various fields, illustrating that curiosity is not limited to people of a single background, race, ethnicity, gender, sexual orientation, or physical ability. We will continue making these films with diverse voices on the creative team with students from the sciences, arts, and humanities, giving the project an inherently broad perspective. Overall, looking to the future, our work will strive to explain the scientific process, and the scientist’s life journey in addition to the scientist’s research to teach and engage the audience in modern science.

\section*{Acknowledgment}


\newcommand{\IMSacktext}{%

The SUPERPOSITION project was partly supported by the National Science Foundation grant NSF DMR-1830704 and by the National Science Foundation, University of Chicago Materials Research and Engineering Center (National Science Foundation MRSEC).  The student funding was provided by UChicago Career Advancement, and College Center for Research and Fellowships, and the equipment was provided by Jonathan Logan Family Media Center, Fire Escape Films, Audio-Visual Services. The authors would like to acknowledge the members of the SUPERPOSITION team that are not co-authors on this paper; Science Advisor: Srivatsan Chakram; Cinematographers: Samuel Audette, Jacob Grayson, Gabrielle Lu, Ugushi Ogonor, Emily Williams; Original Music: Miles Donnelly; Story Editors: Moyosore Abiona, Samuel Audette, Sami Elahi, and Hannah Iafrati; Animation: Sami Elahi;  Series Advising: Moyosore Abiona, Samuel Audette, Jacob Johnson, Anja Krause, Rachel Weathered. The authors also thank STAGE lab’s full \emph{Curiosity} team for developing SUPERPOSITION as well as members of the Schuster\cite{schuster} Lab at the University of Chicago for their willingness to help at every step of the filmmaking process.

}

\IMSdisplayacksection{\IMSacktext}


%
%
\bibliographystyle{IEEEtran}
\bibliography{IEEEreferences}
\end{document}